\documentclass{PoS}

\usepackage{amsmath,amssymb}

\newcommand{\beq}{\begin{equation}}
\newcommand{\eeq}{\end{equation}}
\newcommand{\beqs}{\begin{eqnarray}}
\newcommand{\eeqs}{\end{eqnarray}}
\newcommand{\ZvQQ}{\ensuremath{Z_{V^4_{QQ}}}}
\newcommand{\Zvqq}{\ensuremath{Z_{V^4_{qq}}}}
\newcommand{\MeV}{\ensuremath{\ \textrm{MeV}}}

\title{$B$ and $D$ meson decay constants from 2+1 flavor improved staggered simulations}

\ShortTitle{$B$ and $D$ meson decay constants from 2+1 flavor improved staggered simulations}

\author{%
	\speaker{E.~T.~Neil}$^{,a,\dagger}$,
	Jon~A.~Bailey$^{a,b}$,
	A.~Bazavov$^c$,
	C.~Bernard$^d$,
	C.~M.~Bouchard$^{e,a,g}$,
	C.~DeTar$^h$,
	M.~Di Pierro$^i$,
	A.~X.~El-Khadra$^e$,
	R.~T.~Evans$^e$,
	E.~Freeland$^{e,f}$,
	E.~Gamiz$^{a,j}$,
	Steven~Gottlieb$^k$,
	U.~M.~Heller$^l$,
	J.~E.~Hetrick$^m$,
	R.~Jain$^e$,
	A.~S.~Kronfeld$^a$,
	J.~Laiho$^n$,
	L.~Levkova$^h$,
	P.~B.~Mackenzie$^a$,
	M.~B.~Oktay$^h$,
	J.~N.~Simone$^a$,
	R.~Sugar$^o$,
	D.~Toussaint$^p$,
	and
	R.~S.~Van de Water$^c$ \\ \\
	\llap{$^a$}Fermi National Accelerator Laboratory, Batavia, Illinois, USA \\
	\llap{$^b$}Department of Physics and Astronomy, Seoul National University, Seoul, ROK \\	
	\llap{$^c$}Physics Department, Brookhaven National Laboratory, Upton, NY, USA \\
	\llap{$^d$}Department of Physics, Washington University, St. Louis, Missouri, USA \\
	\llap{$^e$}Physics Department, University of Illinois, Urbana, Illinois, USA \\
	\llap{$^f$}Department of Physics, Benedictine University, Lisle, Illinois, USA \\
	\llap{$^g$}Department of Physics, The Ohio State University, Columbus, OH, USA \\
	\llap{$^h$}Physics Department, University of Utah, Salt Lake City, Utah, USA \\
	\llap{$^i$}School of Computing, DePaul University, Chicago, Illinois, USA \\
	\llap{$^j$}Department of Physics, University of Granada, Granada, Spain \\
	\llap{$^k$}Department of Physics, Indiana University, Bloomington, Indiana, USA \\
	\llap{$^l$}Americal Physical Society, Ridge, New York, USA \\
	\llap{$^m$}Physics Department, University of the Pacific, Stockton, California, USA \\
	\llap{$^n$}SUPA, School of Physics and Astronomy, University of Glasgow, Glasgow, UK \\
	\llap{$^o$}Department of Physics, University of California, Santa Barbara, California, USA \\
	\llap{$^p$}Department of Physics, University of Arizona, Tucson, Arizona, USA \\ \\
	\llap{$^\dagger$}E-mail: \email{eneil@fnal.gov}
}

\author{Fermilab Lattice and MILC Collaborations}

\abstract{We give an update on simulation results for the decay constants $f_B, f_{B_s}, f_D$ and $f_{D_s}$.  These decay constants are important for precision tests of the standard model, in particular entering as inputs to the global CKM unitarity triangle fit.  The results presented here make use of the MILC (2+1)-flavor asqtad ensembles, with heavy quarks incorporated using the clover action with the Fermilab method.  Partially quenched, staggered chiral perturbation theory is used to extract the decay constants at the physical point.  In addition, we give error projections for a new analysis in progress, based on an extended data set.}

\FullConference{XXIX International Symposium on Lattice Field Theory \\
		 July 10-16, 2011\\
		 Squaw Valley, Lake Tahoe, California}

\begin{document}

\section{Introduction\label{sec:intro}}

Within the standard model, the decay of mesons containing heavy quarks (in particular, $B$ and $D$ mesons) into purely leptonic final states provides an important testing ground for a number of theoretical ideas.  Such decays involve both weak and strong interactions simultaneously, so that a complete understanding of the standard model is necessary to describe them.  In particular, the decay width of a charged meson is proportional to both the meson decay constant (determined by strong interactions) and the CKM mixing angle,
\beq
\Gamma(H \rightarrow \ell \nu_\ell) \propto f_H^2 |V_{Qq}|^2.
\eeq

Because this fully leptonic decay has no hadrons in the final state, the meson decay constant $f_P$ can be readily and accurately determined by lattice simulations \cite{hpqcd1,hpqcd2,etmc1,etmc2}.  Such a determination is in fact necessary in order to extract the CKM angles from experimental measurements of these decays, and precise computations of the decay constants could potentially reveal the presence of new physics through tension in the CKM unitarity triangle \cite{arXiv:1008.1593,arXiv:1010.6069,arXiv:1102.3917,Bevan:2010gi}.  In addition, certain leptonic decay channels (e.g. $B_s \rightarrow \mu^+ \mu^-$) are both loop and CKM suppressed in the standard model, and so they may be particularly sensitive to flavor-changing interactions induced by new physics \cite{CDFBs}.

\section{Simulation Details\label{sec:sim}}

We make use of the MILC asqtad-improved staggered gauge configurations, with $2+1$ dynamical quarks in the sea \cite{MILC}.  For the light valence quarks, we make use of the same staggered action, while charm and bottom valence quarks are incorporated using the clover action with the Fermilab interpretation \cite{FermiHQ}.  The particular set of ensembles used to obtain the results presented here, along with the number of configurations and other relevant information, are detailed in Table~\ref{tab:sim}.  Data on the coarsest lattice spacing $a \approx 0.15$ fm are shown in the analysis but are used only for the purpose of estimating the discretization errors; these points are excluded from the final fit used for chiral and continuum extrapolation.

\begin{table}
\begin{center}
\begin{tabular}{|cccccc|}
\hline\hline
$\approx a$ [fm]&$am_h$&$am_l$&$\beta$&$r_1/a$&$N_{\textrm{conf}}$\\
\hline
0.09&0.031&0.0031&7.08&3.75&435\\
&&0.0062&7.09&3.79&557\\
&&0.0124&7.11&3.86&518\\
0.12&0.050&0.005&6.76&2.74&678\\
&&0.007&6.76&2.74&833\\
&&0.010&6.76&2.74&592\\
&&0.020&6.79&2.82&460\\
&&0.030&6.81&2.88&549\\
\textit{0.15}&\textit{0.0484}&\textit{0.0097}&\textit{6.572}&\textit{2.22}&\textit{631}\\
&&\textit{0.0194}&\textit{6.586}&\textit{2.26}&\textit{631}\\
&&\textit{0.0290}&\textit{6.600}&\textit{2.29}&\textit{576}\\
\hline
\end{tabular}
\end{center}
\caption{\label{tab:sim}Table of gauge configurations used for the full analysis to be presented.  The ensembles with $a \approx 0.15$ fm (italics) are excluded from the final chiral/continuum extrapolation, as described in the text.}
\end{table}

\section{Analysis and Fitting\label{sec:fitting}}

The heavy meson decay constant is determined through the overlap of the meson wavefunction $|H\rangle$ with the axial vector current:
\beq
\langle 0 | \mathcal{A}^\mu | H(p) \rangle (M_H)^{-1/2} = i (p^\mu / M_H) (f_H \sqrt{M_H}) \equiv i (p^\mu / M_H) \phi_H.
\eeq
The quantity $\phi_H \equiv f_H \sqrt{M_H}$ is thus proportional to the ground-state amplitude of the two-point function between the axial vector current and a heavy-light pseudoscalar operator $\mathcal{O}$.  We therefore extract $\phi_H$ by fitting this two-point function simultaneously with the two-point pseudoscalar correlator,
\beqs
\Phi_2^{s}(t) &= \frac{1}{4} \sum_{a=1}^4 \langle A_a^{4}{}^\dagger(t,\mathbf{x}) \mathcal{O}_a^{(s)}(0)\rangle, \\
C_2^{s,s'}(t) &= \frac{1}{4} \sum_{a=1}^4 \langle \mathcal{O}_a^{(s)}{}^\dagger(t,\mathbf{x}) \mathcal{O}_a^{(s')}(0)\rangle,
\eeqs
where $a$ is a staggered taste index.  Precise definitions of the interpolating operators $A^{4}$ and $\mathcal{O}$ are given in Ref.~\cite{decaypaper}.

For a correlation function with source type $s$ and sink type $s'$, we fit to the ``factorized" functional form
\beq
C_{ss'}(t) = \sum_{n=0}^{N_X} \left[ A_{s,n} A_{s',n} \left( e^{-E_n t} + e^{-E_n (N_t - t)} \right) - (-1)^t A_{s,n}' A_{s',n}' \left( e^{-E_n' t} +e^{-E_n' (N_t - t)} \right) \right]
\eeq
where $N_X$ denotes the number of excited states included in the fit.  The pseudoscalar source and sink type $s, s'$ can be either point-like or smeared, so that a total of six distinct correlators are available for analysis.  In the results shown here, joint fits are carried out to various combinations of correlators, with Bayesian priors imposed as constraints on the fit parameters.

From the two-point fits, we extract a bare value for the ground-state amplitude between an axial-vector current and pseudoscalar operator, which must then be renormalized in order to obtain the decay constant:
\beq
\phi_{H} = \sqrt{2} Z_{A^4_{Qq}} A_{A^4_{Qq},0}.
\eeq

To compute the heavy-light axial current renormalization constant $Z_{A^4_{Qq}}$, we divide it into flavor-diagonal contributions $Z_{V^4_{qq}}, Z_{V^4_{QQ}}$, which are determined non-perturbatively, and a flavor off-diagonal piece $\rho_{A^4_{Qq}}$ which is computed in lattice perturbation theory \cite{rhoPT}.  Our renormalized result for $\phi_H = f_H \sqrt{M_H}$ is thus given by
\beq
\phi_{H} = \sqrt{2} Z_{A^4_{Qq}} A_{A^4_{Qq},0} = \sqrt{2} (\rho_{A^4_{Qq}} \sqrt{Z_{V^4_{qq}} Z_{V^4_{QQ}}}) A_{A^4_{Qq},0}.
\eeq

We use rooted staggered chiral perturbation theory (rS$\chi$PT) \cite{rSxPT} to extrapolate our results simultaneously to the continuum limit and to the physical light-quark masses.  (Heavy quark masses are tuned non-perturbatively to give physical heavy-light meson masses, so no extrapolation is necessary for them.  Details of the tuning procedure are given in \cite{kappatune}.)  The chiral fit functions incorporate terms describing a number of different effects, including discretization errors, finite-volume corrections, and hyperfine splittings.

\section{Results\label{sec:results}}

\begin{figure}
\begin{center}
\includegraphics[width=95mm]{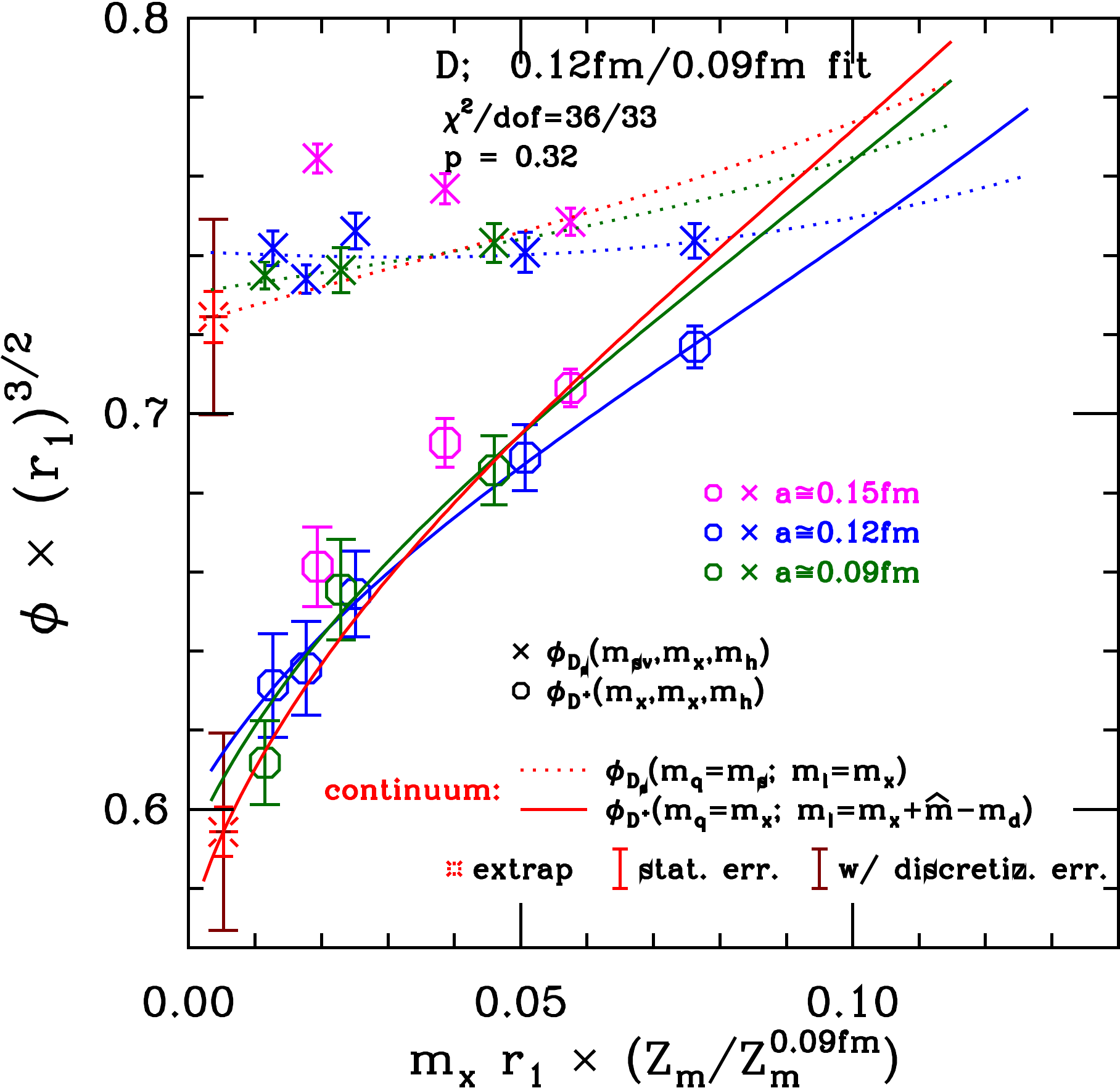}
\caption{Extracted values of $\phi$ and chiral best-fit curves for the $D$ system.  Only points where valence and sea light-quark masses are equal are shown here.  Data from the $a \approx 0.15$ fm ensemble are shown but not included in the fit.  The red curve and symbols show the continuum extrapolation and the continuum/physical point extrapolation, respectively, for both $\phi_{D^+}$ and $\phi_{D_s}$.  For the fully extrapolated points, the inner error bars (bright red) represent statistical errors only, while the outer errors (dark red) include discretization errors. \label{fig:D}}
\end{center}
\end{figure}

\begin{figure}
\begin{center}
\includegraphics[width=95mm]{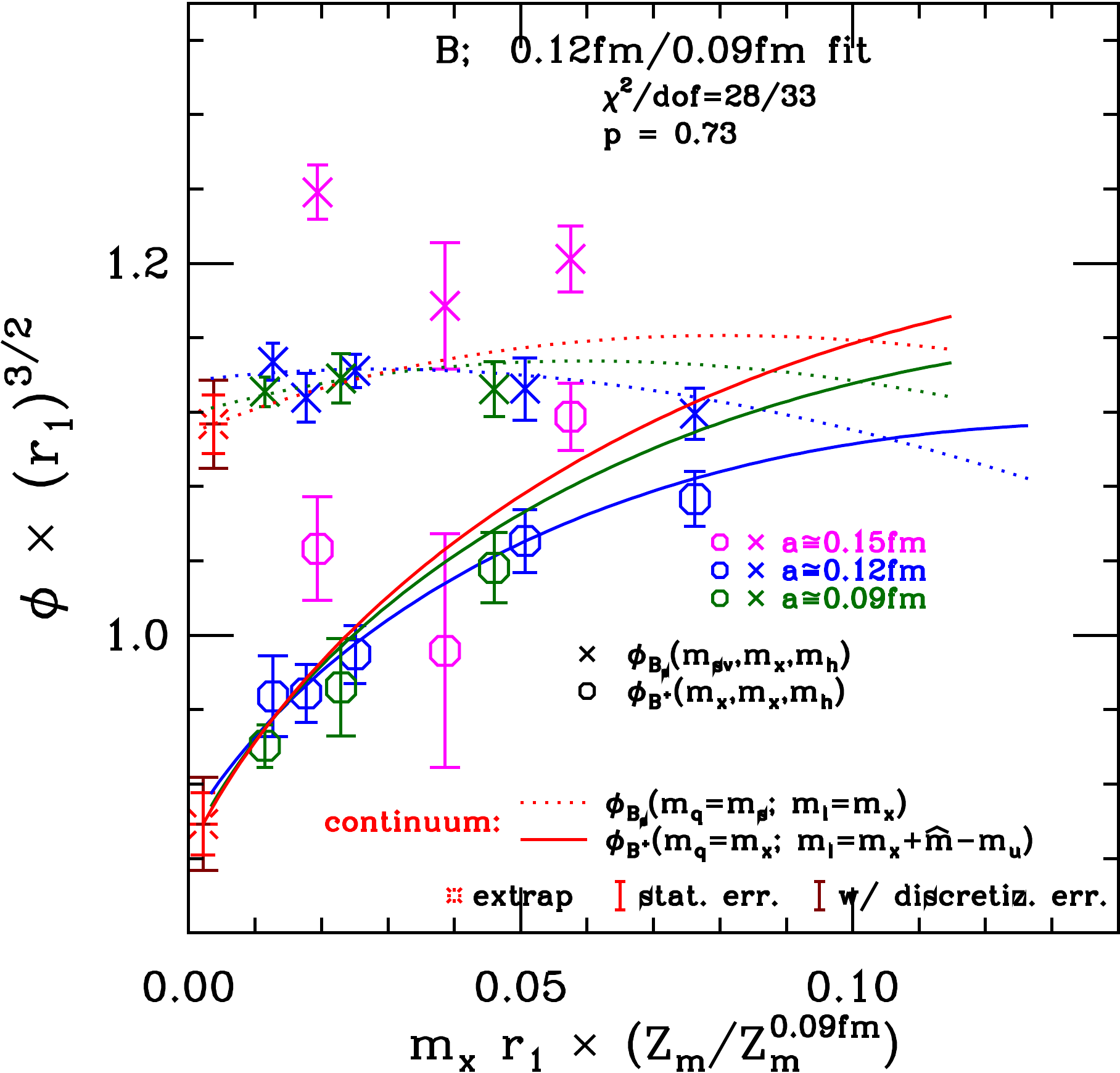}
\caption{$\phi$ values and chiral best-fit curves as in Figure 1, but for the $B$ system.  \label{fig:B}}
\end{center}
\end{figure}

Applying the procedure outlined above, we obtain the values for $\phi$ shown in Figures~\ref{fig:D} and~\ref{fig:B}, renormalized and in units of the standard scale $r_1$.  The chiral best-fit curve to the points is also shown, both explicitly at each lattice spacing and extrapolated to the continuum limit.

Evaluating the continuum best-fit curves at the physical points, we obtain the following values for the decay constants and their ratios:
\begin{align}
f_{B^+} &= 196.9(8.9) \MeV, \\
f_{B_s} &= 242.0(9.5) \MeV, \\
f_{B_s} / f_{B^+} &= 1.229(0.026), \\
f_{D^+} &= 218.9(11.3) \MeV, \\
f_{D_s} &= 260.1(10.8) \MeV, \\
f_{D_s} / f_{D^+} &= 1.188(0.025).
\end{align}
The error bars quoted here include both statistical and systematic sources of error, which are accounted for in a detailed error budget.  A summary of the full error budget for the individual decay constants is given in Table~\ref{tab:err}.  We discuss the error budget further in Section~\protect\ref{sec:outlook} below, but a thorough discussion is beyond the scope of this paper.  Instead, we refer the reader to Ref.~\cite{decaypaper}, which contains a complete discussion of the systematic error analysis, including the full error budget for the decay-constant ratios.

\begin{table}
\begin{center}
	\caption{Error budget for the decay constants, as obtained in Section~\protect\ref{sec:results}.  In addition, projected improvements to the decay constant error budget for the updated analysis in progress (discussed in Section~\protect\ref{sec:outlook}) are shown italicized and in brackets.  \vspace{2mm}\label{tab:err}}
	\begin{tabular}{lllll}
	\hline
	\hline
	Source & $f_{D^+} (\MeV)$ & $f_{D_s} (\MeV)$ & $f_{B^+} (\MeV)$ & $f_{B_s} (\MeV) $  \\
        \hline
        Statistics & 2.3 \emph{[1.1]} & 2.3 \emph{[1.1]} & 3.6 \emph{[1.8]} & 3.4 \emph{[1.7]} \\
        Heavy-quark disc. & 8.2 \emph{[3.6]} & 8.3 \emph{[3.6]} &3.7 \emph{[1.9]} & 3.8 \emph{[2.0]} \\
        Light-quark disc. & 2.9 \emph{[0.7]} & 1.5 \emph{[0.3]} & 2.5 \emph{[0.6]} & 2.1 \emph{[0.5]} \\
        Chiral extrapolation & 3.2 \emph{[1.6]} & 2.2 \emph{[1.1]}  & 2.9 \emph{[1.5]} & 2.8 \emph{[1.4]} \\
        Heavy-quark tuning & 2.8 \emph{[2.0]} & 2.8 \emph{[2.0]}  & 3.9 \emph{[2.4]} & 3.9 \emph{[2.4]} \\
        $\ZvQQ$ and $\Zvqq$ & 2.8 \emph{[1.4]} & 3.4 \emph{[1.7]} & 2.6 \emph{[1.5]} & 3.1 \emph{[1.9]} \\
        $u_0$ adjustment & 1.8 \emph{[0]} & 2.0 \emph{[0]} & 2.5 \emph{[0]} & 2.8 \emph{[0]} \\
        Other sources & 3.8 \emph{[3.8]} & 3.0 \emph{[3.0]}  & 3.5 \emph{[3.5]} & 4.8 \emph{[4.8]} \\
        \hline
        Total \emph{[projected]} error & 11.3 \emph{[6.1]} & 10.8 \emph{[5.6]} & 8.9 \emph{[5.5]} & 9.5 \emph{[6.4]}\\
        \hline
        \hline
        \end{tabular}
\end{center}
\end{table}

\section{Outlook\label{sec:outlook}}

A new analysis following the approach outlined above is currently in progress, based on an expanded set of gauge configurations as shown in Table~\ref{tab:sim2}.  In addition to extending the available simulations to finer lattice spacing and smaller quark mass, the ``new" data set includes large increases in statistics for several ensembles.

Projected improvements in the error budget when the new data set is included are shown alongside the previous error estimates in Table~\ref{tab:err}.  Statistical errors are projected to improve as $\sqrt{N_{\textrm{cfg}}}$, with $N_{\textrm{cfg}}$ the number of gauge configurations available for a given ensemble.  For the various discretization errors, the projected improvements are a result of reducing the smallest lattice spacing available from $a=0.09$ fm to $a=0.045$ fm.  Light-quark discretization errors are estimated to scale as $\mathcal{O}(\alpha_s a^2)$; the heavy-quark discretization errors are estimated using the known functional dependence, which has several terms.  The decrease in the chiral extrapolation error is projected based on the lightest available value of the quark mass in $r_1$ units, $m_x r_1$.  The heavy-quark tuning error is based on a combination of statistical and discretization errors, and is treated as such.  The ``$u_0$ adjustment" error is the result of using different tadpole improvement factors for the valence and sea quarks.  This is rectified in the new data analysis, eliminating the associated error.  Finally, the error estimates for the heavy-quark renormalization factors $\ZvQQ, \Zvqq$ are based on preliminary non-perturbative results for those quantities.

\begin{table}
\begin{center}
\begin{tabular}{|ccccccc|}
\hline\hline
$\approx a$ [fm]&$am_h$&$am_l$&$\beta$&$r_1/a$&$N_{\textrm{conf}}$ (old run)&$N_{\textrm{conf}}$ (new run)\\
\hline
0.045&0.014&0.0028&7.81&7.21&---&800\\
0.06&0.018&0.0018&7.46&5.31&---&825\\
&&0.0025&7.465&5.33&---&800\\
&&0.0036&7.47&5.35&---&631\\
&&0.0072&7.48&5.40&---&591\\
0.09&0.031&0.00155&7.075&3.74&---&790\\
&&0.0031&7.08&3.75&435&577\\
&&0.00465&7.085&3.77&---&983\\
&&0.0062&7.09&3.79&557&1377\\
&&0.0124&7.11&3.86&518&1476\\
0.12&0.050&0.005&6.76&2.74&678&1419\\
&&0.007&6.76&2.74&833&1274\\
&&0.010&6.76&2.74&592&1664\\
&&0.020&6.79&2.82&460&1637\\
&&0.030&6.81&2.88&549&---\\
\textit{0.15}&\textit{0.0484}&\textit{0.0097}&\textit{6.572}&\textit{2.22}&\textit{631}&---\\
&&\textit{0.0194}&\textit{6.586}&\textit{2.26}&\textit{631}&---\\
&&\textit{0.0290}&\textit{6.600}&\textit{2.29}&\textit{576}&---\\
\hline
\end{tabular}
\end{center}
\caption{\label{tab:sim2} Table of gauge configurations for the updated analysis in progress.  Only configurations labelled ``old run" were used to obtain the results presented in Section 4.  The analysis in progress will make use of all configurations from both runs combined.}
\end{table}

\section*{Acknowledgments}
Computations for this work were carried out with resources provided by
the USQCD Collaboration, the Argonne Leadership Computing Facility,
the National Energy Research Scientific Computing Center, and the 
Los Alamos National Laboratory, which are funded by the Office of Science of the
U.S. Department of Energy; and with resources provided by the National Institute
for Computational Science, the Pittsburgh Supercomputer Center, the San Diego
Supercomputer Center, and the Texas Advanced Computing Center, which are funded
through the National Science Foundation's Teragrid/XSEDE Program.
This work was supported in part by the U.S. Department of Energy under 
Grants No.~DE-FC02-06ER41446 (C.D., L.L., M.B.O.),
No.~DE-FG02-91ER40661 (S.G.), 
No.~DE-FG02-91ER40677 (C.M.B, R.T.E., E.D.F., E.G., R.J., A.X.K.),
No.~DE-FG02-91ER40628 (C.B),
No.~DE-FG02-04ER-41298 (D.T.); 
by the National Science Foundation under Grants 
No.~PHY-0555243, No.~PHY-0757333, No.~PHY-0703296 (C.D., L.L., M.B.O.), 
No.~PHY-0757035 (R.S.), and
No.~PHY-0704171 (J.E.H.);
by the URA Visiting Scholars' program (C.M.B., R.T.E., E.G., M.B.O.);
and by the Fermilab Fellowship in Theoretical Physics (C.M.B.).
This manuscript has been co-authored by employees of Brookhaven Science
Associates, LLC, under Contract No. DE-AC02-98CH10886 with the 
U.S. Department of Energy.
Fermilab is operated by Fermi Research Alliance, LLC, under Contract
No.~DE-AC02-07CH11359 with the United States Department of Energy.


\begin{thebibliography}{99}

  
\bibitem{hpqcd1} 
  C.~McNeile, C.~T.~H.~Davies, E.~Follana, K.~Hornbostel and G.~P.~Lepage,
  arXiv:1110.4510 [hep-lat].
  
\bibitem{hpqcd2} 
  C.~T.~H.~Davies {\it et al.},
  Phys.\ Rev.\ D\ {\bf 82}, 114504  (2010)
  [arXiv:1008.4018 [hep-lat]].
  
\bibitem{etmc1} 
  P.~Dimopoulos {\it et al.} [ETM Collaboration],
  arXiv:1107.1441 [hep-lat].
  
\bibitem{etmc2} 
  F.~Farchioni {\it et al.},
  PoSLATTICE\ {\bf 2010}, 128  (2010)
  [arXiv:1012.0200 [hep-lat]].

\bibitem{arXiv:1008.1593} 
  A.~Lenz {\it et al.} [CKMfitter Collaboration],
  Phys.\ Rev.\ D\ {\bf 83}, 036004  (2011)
  [arXiv:1008.1593 [hep-ph]].

\bibitem{arXiv:1010.6069} 
  E.~Lunghi and A.~Soni,
  Phys.\ Lett.\ B\ {\bf 697}, 323  (2011)
  [arXiv:1010.6069 [hep-ph]].

\bibitem{arXiv:1102.3917} 
  J.~Laiho, E.~Lunghi and R.~Van De Water,
  PoSFPCP\ {\bf 2010}, 040  (2010)
  [arXiv:1102.3917 [hep-ph]].
  
\bibitem{Bevan:2010gi}
  A.~J.~Bevan {\it et al.}  [UTfit Collaboration],
  PoS {\bf ICHEP2010}, 270 (2010)
  [arXiv:1010.5089 [hep-ph]].

\bibitem{CDFBs} 
  T.~Aaltonen {\it et al.} [CDF Collaboration],
  Phys.\ Rev.\ Lett.\ \ {\bf 107}, 191801  (2011)
  [arXiv:1107.2304 [hep-ex]].
  
\bibitem{MILC} 
  A.~Bazavov {\it et al.},
  Rev.\ Mod.\ Phys.\ \ {\bf 82}, 1349  (2010)
  [arXiv:0903.3598 [hep-lat]].
  
  \bibitem{FermiHQ} 
  A.~X.~El-Khadra, A.~S.~Kronfeld and P.~B.~Mackenzie,
  Phys.\ Rev.\ D\ {\bf 55}, 3933  (1997)
  [hep-lat/9604004].

 \bibitem{decaypaper}
 A.~Bazavov {\it et al.} [Fermilab Lattice and MILC Collaborations],
 ``$B$- and $D$-meson decay constants from three-flavor lattice QCD,"
  [arXiv:1112.3051 [hep-lat]].

\bibitem{rhoPT} 
  A.~X.~El-Khadra {\it et al.},
  PoSLAT\ {\bf 2007}, 242  (2007)
  [arXiv:0710.1437 [hep-lat]].
  
\bibitem{rSxPT} 
  C.~Aubin and C.~Bernard,
  Phys.\ Rev.\ D\ {\bf 73}, 014515  (2006)
  [hep-lat/0510088].

\bibitem{kappatune} 
  C.~Bernard {\it et al.} [Fermilab Lattice and MILC Collaborations],
  Phys.\ Rev.\ D\ {\bf 83}, 034503  (2011)
  [arXiv:1003.1937 [hep-lat]].
  


\end{thebibliography}
\end{document}